\RequirePackage{fix-cm}

\documentclass[smallextended]{svjour3}       
\smartqed  

\usepackage{array}
\usepackage{mathptmx}
\usepackage{graphicx}
\usepackage{listings}
\usepackage{verbatimbox}
\usepackage{subfig}
\usepackage{paralist}
\usepackage{xspace}
\usepackage{booktabs}
\usepackage{color}
\usepackage{setspace}
\usepackage{url}
\usepackage[numbers, sort&compress]{natbib}
\usepackage[colorlinks,bookmarksopen,bookmarksnumbered,citecolor=red,urlcolor=red]{hyperref}
\usepackage{flushend}
\usepackage{subfig}
\usepackage{multirow}
\usepackage{color}
\usepackage{float}
\usepackage{amsmath}
\usepackage{amssymb}
\usepackage{fancybox}
\usepackage{rotating}
\usepackage{tablefootnote}
\usepackage[para,online,flushleft]{threeparttable}
\usepackage{multirow}
\usepackage{hhline}
\usepackage{enumitem}
\usepackage[dvipsnames]{xcolor}
\colorlet{LightRubineRed}{RubineRed!70!}

\newcommand{\hypobox}[1]{\begin{center}%
	\noindent\thicklines\setlength{\fboxsep}{7pt}%
	\cornersize{0}\Ovalbox{\begin{minipage}{0.9\textwidth}%
	\vspace{-0.1cm}
	\textit{#1}
	\vspace{-0.1cm}
	\end{minipage}} \end{center}}

\newcommand{\code}[1]{{\texttt{#1}}}

\definecolor{mygray}{gray}{0.4}
\newfloat{Code}{H}{myc}

\begin{document}

\title{An Empirical Study of the Characteristics of Popular Minecraft Mods}
\author{
%
%
\alignauthor
Daniel~Lee\\
       \affaddr{Software Analysis and Intelligence Lab (SAIL)}\\
       \affaddr{Queen's University}\\
       \affaddr{Kingston, ON, Canada}\\
       \email{dlee@cs.queensu.ca}
\alignauthor
Gopi~Krishnan~Rajbahadur\\
       \affaddr{Software Analysis and Intelligence Lab (SAIL)}\\
       \affaddr{Queen's University}\\
       \affaddr{Kingston, ON, Canada}\\
       \email{krishnan@cs.queensu.ca}
\alignauthor
Dayi~Lin\\
       \affaddr{Software Analysis and Intelligence Lab (SAIL)}\\
       \affaddr{Queen's University}\\
       \affaddr{Kingston, ON, Canada}\\
       \email{dayi.lin@cs.queensu.ca} 
\alignauthor       
Mohammed~Sayagh\\
       \affaddr{Software Analysis and Intelligence Lab (SAIL)}\\
       \affaddr{Queen's University}\\
       \affaddr{Kingston, ON, Canada}\\
       \email{msayagh@cs.queensu.ca}
\alignauthor
Cor-Paul~Bezemer\\
       \affaddr{Analytics of Software, Games and Repository Data (ASGAARD) Lab}\\
       \affaddr{University of Alberta}\\
       \affaddr{Edmonton, AB, Canada}\\
       \email{bezemer@ualberta.ca}
\alignauthor
Ahmed~E.~Hassan\\
       \affaddr{Software Analysis and Intelligence Lab (SAIL)}\\
       \affaddr{Queen's University}\\
       \affaddr{Kingston, ON, Canada}\\
       \email{ahmed@cs.queensu.ca}
}

\author{Daniel~Lee \and  Gopi~Krishnan~Rajbahadur \and Dayi~Lin \and Mohammed~Sayagh \and Cor-Paul~Bezemer \and Ahmed~E.~Hassan }


\institute{Daniel Lee \and Gopi Krishnan Rajbahadur \and Dayi Lin \and Mohammed Sayagh \and Ahmed E. Hassan  \at
              Software Analysis and Intelligence Lab (SAIL)\\
              Queen's University\\
              Kingston, ON, Canada \\
              \email{\{dlee, krishnan, dayi.lin, msayagh, ahmed\}@cs.queensu.ca}  \\ \\
Cor-Paul Bezemer \at
              Analytics of Software, Games and Repository Data (ASGAARD) Lab\\
              University of Alberta\\
              Edmonton, AB, Canada \\
              \email{bezemer@ualberta.ca}  \\
}

\date{Received: date / Accepted: date}

\maketitle

\begin{abstract} 
It is becoming increasingly difficult for game developers to manage the cost of developing a game, while meeting the high expectations of gamers. One way to balance the increasing gamer expectation and development stress is to build an active modding community around the game. 
There exist several examples of games with an extremely active and successful modding community, with the Minecraft game being one of the most notable ones.

This paper reports on an empirical study of 1,114 popular and 1,114 unpopular Minecraft mods from the CurseForge mod distribution platform, one of the largest distribution platforms for Minecraft mods. We analyzed the relationship between 33 features across 5 dimensions of mod characteristics and the popularity of mods (i.e., mod category, mod documentation, environmental context of the mod, remuneration for the mod, and community contribution for the mod), to understand the characteristics of popular Minecraft mods. We firstly verify that the studied dimensions have significant explanatory power in distinguishing the popularity of the studied mods. Then we evaluated the contribution of each of the 33 features across the 5 dimensions. We observed that popular mods tend to have a high quality description and promote community contribution.
\end{abstract}


\section{Introduction}
\label{sec:intro}

The team size, cost and complexity in game development can grow exponentially as the user requirements increase~\cite{howmuchdoesitcost}. Thus, it has become challenging to develop a successful game, and game developers are constantly under an immense amount of stress~\cite{humancostofrd}.

One approach to balance the increasing gamer expectation and development stress is to build an active modding community around the game. Skyrim and Minecraft are examples of games that have been successful in building active modding communities~\cite{hackman2014modders, zorn2013exploring} to increase the longevity of the games. For example, the Skyrim game still has a median of 86 new mods released per day 8 years after its initial game release in 2011, along with more than 514M total unique downloads of mods~\cite{nexusmodsplatform}. Prior work also shows that an active modding community can contribute to the increased sales of the original game~\cite{placingvalueoncommunitycocreations}.

There are two key components of an active modding community of a game: the active development of mods, and the active adoption of mods by gamers. In our prior work, we looked at how game developers can help maintain the active development of mods, and observed that games from developers with a consistent modding support within the same or different game franchises, were associated with faster releases of mods~\cite{dannexus2019}. In this paper, we identify the characteristics that distinguish popular mods from unpopular ones. To do so, we study 33 characteristics along 5 dimensions of 1,114 popular and 1,114 unpopular mods for the Minecraft game from the CurseForge mod distribution platform -- one of the largest distribution platforms for Minecraft mods. We focus on the mods from the Minecraft game because it has one of the largest and most active modding communities~\cite{activemoddingcommunity}. In particular, we answer the following two research questions (RQs):

\begin{description}
    \item [\textbf{RQ1:}] \textbf{Do our studied dimensions have enough explanatory power to distinguish popular mods from unpopular ones?} 
	
    \textit{Motivation: }The goal of this research question is to investigate how well each studied dimension of characteristics (i.e., features) of mods can individually distinguish the popular mods from unpopular ones. We also investigate how well all the studied dimensions together can distinguish popular mods from unpopular ones. Prior work~\cite{mobilechar} used similar dimensions to identify the characteristics that distinguish mobile apps with high ratings from the ones with low ratings. The results of this research question lay the foundation for further investigations of the characteristics of popular mods.
	
    \textit{Findings: }We observed that each studied dimension of characteristics of a mod has significant explanatory power in distinguishing popular from unpopular mods. Among the studied dimensions, the community contribution for the mod dimension has the largest explanatory power. However, our combined model which uses all the features across the five dimensions outperforms the best model using an individual dimension by 10\% (median).

    \item [\textbf{RQ2:}] \textbf{What features best characterize a popular mod?} 
	
    \textit{Motivation: }The goal of this research question is to investigate which features of mods can best characterize popular mods. The results of RQ1 show that the studied features have a strong explanatory power for the popularity of a mod. In this RQ, we further investigate the characteristics of popular mods at a granular level. 
	
	\textit{Findings: }We observed that 18 of the 33 (54.5\%) studied features help in distinguishing popular mods from unpopular ones. 
	Simplifying the mod development is positively correlated with mod popularity. 
	In addition, popular mods tend to promote community contribution with a source code repository URL and an issue tracking URL, and have a richer mod description. 
\end{description}

The remainder of the paper is outlined as follows. Section~\ref{sec:bg} gives background information about the Minecraft game and the CurseForge mod distribution platform. Section~\ref{sec:related} gives an overview of related work. Section~\ref{sec:method} discusses our methodology. Sections~\ref{sec:results} discusses the results of our empirical study. 
Section~\ref{sec:threats} outlines threats to the validity of our findings. Section~\ref{sec:conclusion} concludes our study.

\section{Background}
\label{sec:bg}
This section provides a brief overview of the Minecraft game and the CurseForge mod distribution platform.

\begin{table}[!t]
\caption{An overview of Minecraft mod distribution platforms}
\centering

\begin{tabular}{lr}
\toprule

\textbf{\begin{tabular}[l]{@{}l@{}}Minecraft mod \\distribution platform\end{tabular}} & \textbf{\# of mods}\\
\midrule
CurseForge\textsuperscript{1}& 12,710\\ 
Planet Minecraft\textsuperscript{2}& 9,159 \\
Minecraft Six\textsuperscript{3}& 3,880\\
Minecraft Mods\textsuperscript{4}& 532 \\

\bottomrule
\multicolumn{2}{l}{\small{$^1$ \url{https://minecraft.curseforge.com}}} \\
\multicolumn{2}{l}{\small{$^2$ \url{https://www.planetminecraft.com}}} \\
\multicolumn{2}{l}{\small{$^3$ \url{http://minecraftsix.com}}} \\
\multicolumn{2}{l}{\small{$^4$ \url{https://www.minecraftmods.com}}} \\
\end{tabular}
\label{tab:mod_platform_overview}
\end{table}

\subsection{The Minecraft Game}

The Minecraft game is an open-ended 3D sandbox game, initially developed in the Java programming language, where gamers can use various resources (e.g., blocks) to create their own worlds~\cite{minecraft}. Developed by the Mojang\footnote{\url{https://mojang.com/}} game studio, the Minecraft game is one of the best selling video games of all time in 2019, with over 176 million copies sold since its release in 2011~\cite{mcbestselling2019}. Mods are considered one of the most popular aspects of the Minecraft game, and are credited for the great success of the game~\cite{bestmcmods,modmakesmcpop1, modmakesmcpop2}.

\subsection{The CurseForge Mod Distribution Platform}

\textbf{Minecraft mods on CurseForge. }The CurseForge mod distribution platform hosts one of the largest online Minecraft mod repositories with more than 12,000 downloadable mods~\cite{cflargestmcrepo}. Table~\ref{tab:mod_platform_overview} shows a comparison of the CurseForge mod distribution platform to other Minecraft mod distribution platforms with respect to the number of mods. The CurseForge mod distribution platform provides a dedicated page for each mod. The dedicated page contains detailed information about a mod including contributors, releases, and dependencies, while categorizing the mod under at least one mod category. Furthermore, mod developers can provide their Paypal\footnote{\url{https://www.paypal.com/}} or Patreon\footnote{\url{https://www.patreon.com/}} donation URLs on their mod's page. Patreon is a crowdfunding platform where content creators such as mod developers can promote themselves, and receive monthly donations.

\textbf{Mod contributors on CurseForge. }A mod on the CurseForge mod distribution platform can have multiple contributors, and each contributor is assigned a role for the mod (i.e., artist, author, contributor, documenter, former author, maintainer, mascot, owner, tester, ticket manager, or translator). There can be multiple contributors of a mod with the same role, except for the ``owner'' role which is only assigned to the user that creates the mod on the platform. Unfortunately, the CurseForge mod distribution platform does not provide any official definition for the roles. 
Furthermore, we observed that the number of mod developers in a mod does not always accurately represent the actual number of contributors. For example, the \textit{Fossils and Archeology Revival} mod\footnote{\url{https://minecraft.curseforge.com/projects/fossils}} shows 10 mod developers on the CurseForge page, but the mod has 17 contributors on Github. 
Hence, we do not use the mod developer roles or the number of mod developers in our study.

\textbf{Mod releases and dependencies on CurseForge. } The dedicated page of each mod on the CurseForge mod distribution platform lists the mod releases with corresponding upload dates and supported Minecraft, Java, and Bukkit\footnote{Bukkit is a Minecraft Server mod that helps in the running and modification of a Minecraft server. See \url{https://bukkit.org/pages/about-us/} for more details.} versions. In addition, the dependencies for each release are also listed on a mod's page. The CurseForge mod distribution platform supports the declaration of several types of dependencies of a mod release, including ``incompatible'', ``tool'', ``required'', ``embedded library'', and ``optional dependencies''.

\section{Related Work}
\label{sec:related}

This section discusses prior studies that are related to our study. We discuss related work on (1)~empirical studies of game mods,
(2)~games and software engineering, 
(3)~studies of the Minecraft game, and (4)~mining online software distribution platforms.

\subsection{Empirical Studies of Game Mods}

Several prior studies studied the modding community to identify and analyze the relationship between mod developers and the game industry, yielding insights on collaborative practices and strategies, as well as capturing the value of mods~\cite{digitalconsumernetworks, profitingfrominnovativeuser, industriallogic}. A few prior studies mined data from the Nexus Mods distribution platform to quantitatively study the motivation behind mod developers based on the users' expectations, and to understand how to build and maintain an active modding community~\cite{populargamemods, dannexus2019}. Particularly, Dey~et~al.~\cite{populargamemods} study the meta data available for popular and unpopular mods of six famous PC games across several popular online mod distribution platforms to investigate the motivations of mod developers. They find that user demands and the content created by the mod developers correlate very weakly and suggest that more effort needs to undertaken to bridge this gap. Furthermore, similar to our study they also seek to investigate what features make a mod popular. However, they consider only the general tags associated with a given mod and they do it across multiple games without any consideration to the game-specific characteristics.

Additionally, Poretski and Arazy~\cite{placingvalueoncommunitycocreations} conducted an empirical study on 45 games from the Nexus Mods distribution platform and observed that mods increased the sales of the original game. Targett~et~al.~\cite{targett2012study} empirically studied user-interface mods of the World of Warcraft\footnote{\url{https://worldofwarcraft.com/en-us/}} game to gather insights on how mods contribute to the World of Warcraft game and its modding community. They observed that modifications helped the interface of video games meet the needs of users, since every user has their own ideal interface. 

Similarly, Wu~et~al.~\cite{wu2016video} studied popular Reddit threads on Minecraft mod discussions to uncover the learnt knowledge by Minecraft modders. They assert that these threads contain vast peer-generated knowledge on how to create artifacts in the Minecraft environment. Levitt~\cite{leavitt2013source} studied the evolution of the creative process around the creation of Minecraft mods. Additionally, several studies~\cite{lane2017minecraft,nguyen2016minecraft} investigated Minecraft mods and their role in enhancing individual creativity and general interest in the field of Science, Technology, Engineering and Mathematics (STEM). They found that modding in the context of the Minecraft game positively influenced both of these aforementioned aspects. Beggs~\cite{beggs2012minecraft} studied how the dynamics between producers and consumers within the game industry are impacted by modding. They did so by studying Minecraft mods. Beggs observed that Minecraft modders in total spend close to 3 million hours weekly creating and maintaining mods. Furthermore, they also noted that the modding culture pushes game consumers into generally preferring games that allow modding.

Different from the aforementioned studies, we study the characteristics that distinguish popular mods from unpopular ones specific to a particular game (Minecraft) in order to better understand the characteristics of popular mods.

\subsection{Games and Software Engineering}
Several studies investigated open source game projects to relate them to software engineering aspects~\cite{ahmed2017open, gamevsnongame}. For instance, Pascerella~et~al.\cite{gamevsnongame} investigated how the developers contribute to video games in an open source setting. A few studies analyzed the development of the authors' own video games~\cite{graham2006toward, kohler2012feedback}, while Guana et al.~\cite{guana2015building} studied the development of their own game engine. In particular, Guana et al.~\cite{guana2015building} outline how game development is more complicated than traditional software development and presents a model-driven approach to simplify the development of game engines. B{\'e}cares et al.~\cite{becares2017approach} investigated the gameplay of the \textit{Time and Space} game and outlined an approach to automate the game tests. 

A few prior studies studied the videos of game-related bugs~\cite{lewis2010went}. Notably, Lin et al.~\cite{linidentifying} identified gameplay videos that showcase game bugs, as na\"ive methods such as keyword search is inaccurate. They proposed a random forest classifier that out-performs other classifiers (i.e., logistic regression and neural network), and provides a precision that is 43\% higher than the na\"ive keyword search approach. Furthermore, several studies~\cite{lewis2011whats,asurveyonactualsoftwareengineeringprocesses,Washburn2016} have been conducted on the postmortems of games based on articles/magazines to draw insights on the do's and dont's of game development.

Ampatzoglou and Stamelos~\cite{ampatzoglou2010software} provided researchers with a systemic review on available literature. In addition, Scacchi and Cooper~\cite{scacchi2015research} extensively analyzed the software engineering literature of games.

Rather than investigating the software engineering aspect of the original game, in this paper we conduct an empirical study by mining the software engineering aspects of game mods that are available in the CurseForge platform.

\subsection{Studies of the Minecraft Game}

Several prior studies have examined the Minecraft game for pedagogical uses~\cite{nebel2016mining,stone2019online, lenig2018minecrafting,al2014design,bayliss2012teaching,bebbington2014case,brand2013crafting,duncan2011minecraft,ekaputra2013minecraft,hanghoj2014redesigning,petrov2014using,short2012teaching,siko2011beyond,zorn2013exploring}. In addition, Nebel et al.~\cite{nebel2016mining} conducted an extensive literature review on the usage of the Minecraft game in education. A few prior studies primarily focused on using the Minecraft game to study the players of the game~\cite{canossa2013give,muller2015statistical,quiring2015voxel}. Furthermore, a few prior studies primarily focused on using the Minecraft game to streamline the development of software~\cite{balogh2013codemetrpolis, saito2014minecraft}.

In our study, we analyze Minecraft mods to provide an empirical understanding of the characteristics of popular mods.

\subsection{Mining Online Software Distribution Platforms}

Mining online software distribution platforms to provide useful information and insights about the popularity of software has been a fundamental part of software engineering research. We present a brief summary of how mining online software distribution platforms has been carried out in the context of traditional software, games and mobile apps.

\textit{Traditional software.} GitHub is one of the most popular online code hosting distribution platforms for traditional software. Several prior studies investigated the popularity of software projects in GitHub to provide insights to software developers~\cite{zhu2014patterns,blincoe2016understanding,kalliamvakou2014promises,borges2016predicting,borges2016understanding,borges2018s}. For example, Borges~et~al.~\cite{borges2016understanding} outline how a GitHub repository gathers popularity over time. In addition, Borges et al. outline the characteristics of successful GitHub repositories for other software developers to mimic. Similarly, Zhu~et~al.~\cite{zhu2014patterns} suggest that better folder organizational practices lead to better project popularity in GitHub.

\textit{Mobile apps.} Many prior studies investigated features that impact the success of a mobile app by mining data from mobile app stores to provide useful guidelines to mobile app developers~\cite{appchurnmobilepop,mobilechar,bavota2014impact,linares2013api,taba2014exploratory,chia2012app}. For example, Tian~et~al.~\cite{mobilechar} studied the differences between popular and unpopular mobile apps and found that popular apps generally have more complex code and better exploit the latest features of the target Android SDK (Software Development Kit). Taba~et~al.~\cite{taba2014exploratory} studied how the complexity of the UI of a mobile app affects its popularity and provided guidelines to developers on the amount of UI complexity they should strive for in order to keep their users happy. Similarly, Bavota~et~al.~\cite{bavota2014impact} and Linares-V{\'a}squez~et~al.~\cite{linares2013api} studied the characteristics of the APIs used by popular and unpopular apps and recommended developers to use less defect-prone and change-prone APIs to ensure the popularity of their mobile apps.

\textit{Games.} Prior studies that mine data from online game distribution platforms primarily focused on extrapolating useful insights for game developers from platforms such as Steam~\cite{theplaytimeprinciple,cheatinginonlinegamesasocialnetwork,lin2018gr}. For example, Lin~et~al.~\cite{urgentupdates} studied urgent updates on the Steam platform and observed several update patterns to help developers avoid undesirable updates. Lin~et~al.~\cite{lin2017eag} also studied the early access model on the Steam platform and suggested that game developers use the early access model to elicit early feedback and gather more positive feedback. Cheung~et~al.~\cite{cheung2014first} investigated over 200 Xbox 360 game reviews to understand how the first hour of gameplay engages new players. Similarly, Ahn~et~al.~\cite{ahn2017makes} analyzed game reviews between popular and unpopular games on the Steam platform to better understand the characteristics of popular Steam games, and offered guidance to game developers on how to make their game popular.

Though many studies mined various software repositories and provided insights to developers, these insights do not directly translate to mod developers as software such as mobile apps and games are developed from the ground-up for the consumption of users. In contrast, game mods are software that was built to enhance, extend or provide (new) features to an existing game in a meaningful way by hacking the source code of the original or through official APIs. Several prior studies~\cite{videogamedevelopmentdifferentfromsoftwaredev,gamevsnongame,petrillo2009went,petrillo2008houston} show that video game development is starkly different from other types of software development. Therefore, by extension, we expect game mod development (which is a subset of game development) to be different from mobile app and video games development. For instance, consider these two studies by Tian~et~al.~\cite{mobilechar} and Ahn~et~al.~\cite{ahn2017makes}. Both studies examine the characteristics of popular mobile apps and video games by mining the Google Play store and the Steam platform respectively to provide insights to mobile app and video game developers. For the mobile app developers, Tian~et~al.~\cite{mobilechar} suggest that size of the app, number of promotional images and the target SDK are the three key elements that are associated with the popularity of a mobile app. In contrast, Ahn~et~al.~\cite{ahn2017makes} recommend developers to improve the gameplay, the challenge and the motivational aspects and emotional connect of the video game while lowering the price and improving the game’s storyline. However, different from both of these studies, from studying the CurseForge platform we find that popular mods are likely to have a better mod description, ease other mod development and welcome community contributions. Such a result further signifies that game mods are different from other types of software.

Hence, the findings and recommendations for mobile developers, game developers and traditional software developers to ensure the popularity of their software as prescribed by prior studies cannot be directly transferred to game mod developers. Therefore, a study such as ours is pivotal in understanding the characteristics of popular mods. We envision future studies to build on our work in order to help developers improve the popularity of their mods.

We did however conduct our study in the same vein as the aforementioned studies by mining the CurseForge mod distribution platform to gain an empirical understanding of the characteristics of popular mods. To the best of our knowledge, the study by Dey~et~al.~\cite{populargamemods} is the only other study that mines online mod distribution platforms to study the characteristics of popular mods. However, they focus only on the tags that are provided for the mods on the distribution platforms and do not endeavour to provide insights to mod developers.

We study the characteristics of popular and unpopular mods specific to a particular game (Minecraft) to better understand what characterizes popular mods. These characteristics can be further explored by future work to assist mod developers in improving the quality of their mods. Furthermore, we are the first to conduct a statistically rigorous analysis on 33 features collected across 5 dimensions to generate insights for mod developers.

\section{Methodology}
\label{sec:method}

\begin{figure*}[!t]
\center
 \includegraphics[width=1\textwidth]{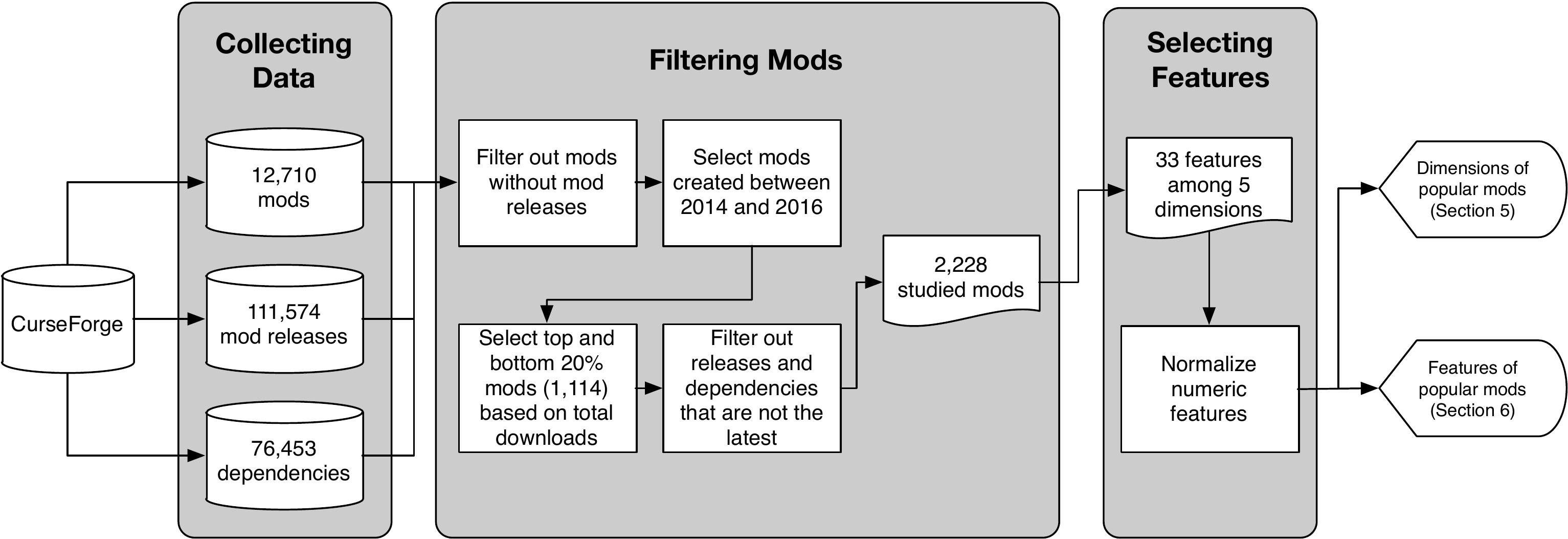}
 \caption{An overview of our data collection.}
 \label{fig:methodology_overview}
\end{figure*}

This section discusses the methodology of our empirical study of the characteristics of popular and unpopular Minecraft mods. 
Figure~\ref{fig:methodology_overview} gives an overview of our methodology.
\subsection{Collecting Data}
We collected the dataset for our study from the CurseForge mod distribution platform on June~6, 2019, using a customized crawler. Table~\ref{tab:mod_dataset_overview} shows an overview our Minecraft mod dataset.

\noindent\textbf{Collecting Mods.} We collected the information of 12,710 mods. In particular, we collected the name, categories, number of total comments, source code URL, issue tracking URL, Paypal URL, and Patreon URL for each mod.

\noindent\textbf{Collecting Mod Releases.} We collected the information of 111,574 releases across all mods. In particular, we collected the type, upload date, size, number of downloads, and supported Minecraft, Java, and Bukkit versions for each mod release.

\noindent\textbf{Collecting Dependencies.} We collected 76,453 mod dependencies across all mod releases. In particular, we collected the type, mods, and the direction for each dependency.

\begin{table} [!t]
	\centering
	\caption{An overview of the CurseForge mod distribution platform dataset.} 
	\label{tab:mod_dataset_overview}
	\begin{tabular} {lr}
	\toprule
	\bfseries Number of total mods & 12,710 \\
 \bfseries Number of studied mods & 2,228 \\
	\addlinespace
	\bfseries Number of studied dimensions & 5\\
	\bfseries Number of studied features & 33\\
	\addlinespace

	\bfseries Number of total mod releases & 111,574\\

	\bfseries Number of total mod dependencies & 76,453 \\

	\bottomrule
	\end{tabular}
\end{table}

\subsection{Filtering Mods}

To ensure the quality of the studied mods, we first removed 295 inactive mods that have no mod releases. Then, we removed 6,845 mods that were created before 2014 or after 2016 to ensure the studied mods all have an equal chance to obtain a high number of downloads. For the remaining 5,570 mods, we selected the top and bottom 20\% of the mods based on their total number of downloads for our study. We consider the top 20\% of mods (1,114 mods) as popular mods, and the bottom 20\% of mods (1,114 mods) as unpopular mods based on their total number of downloads. Hence the claims that are made about a mod being (un)popular are about the likelihood of the mod belonging to the most/least popular group of mods.

\begin{figure*}[!t]
\center
 \includegraphics[width=1\textwidth]{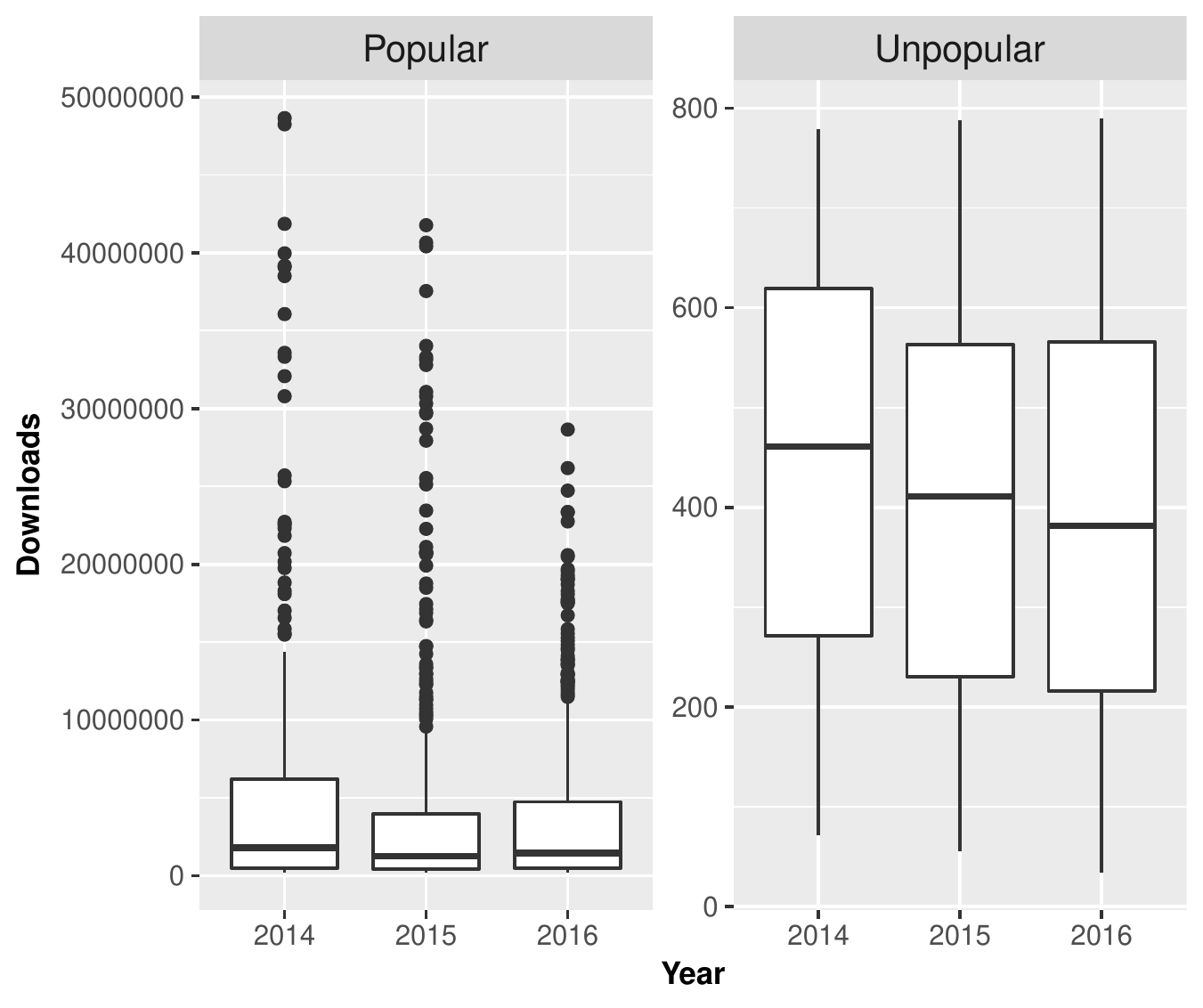}
 \caption{Distribution of the number of downloads that are received by popular and unpopular mods that are created in 2014, 2015 and 2016.}
 \label{fig:popular_unpopular}
\end{figure*}

We do not take into account the lifetime of a mod (despite some mods being created in 2014 and some mods being created in 2016) when separating the mods into popular and unpopular groups. We do so as the number of median downloads across the studied years for mods in the popular and unpopular groups remains relatively consistent as we can observe from Figure~\ref{fig:popular_unpopular}. Furthermore, we observed that the number of popular mods that were created each year in the studied period also remains consistent. More specifically, among the 1,114 popular mods, 279 were created in 2014, and 415 and 418 mods were created in 2015 and 2016 respectively. In total, we studied 2,228 mods. Our selection approach is similar to prior study~\cite{mobilechar} which selected the highest and lowest rated mobile apps for study.

We choose to study the number of downloads as a proxy for the popularity of a mod, as this number acts as a good indicator of the needs for the provided features/alterations by the mod within the Minecraft community. Furthermore, a mod becoming popular in an online platform like CurseForge is pivotal for the mod developers. For instance, as Postigo~et~al.~\cite{ofmodandmodders} outline, mod developers want their mods to be popular as being known in the modding community may open up potentially lucrative job opportunities. Finally, identifying features that affect the popularity of software in online distribution platforms is widely regarded as an important software engineering challenge~\cite{nagappan2016future}. This importance is for example demonstrated by the many software engineering studies that examine the characteristics of popular mobile apps in app stores (e.g.,\cite{harman2012app,mobilechar,bavota2014impact,linares2013api}).

For each of the 2,228 mods, we used the information of the mod's latest release and dependencies in our study.

\subsection{Selecting Features}
Starting from prior work on the popularity of mobile apps~\cite{mobilechar} and our own intuition, we defined 5 dimensions that might be associated with the popularity of mods (i.e., mod category, mod documentation, environmental context of the mod, remuneration for the mod, and community contribution for the mod). Then, we define for each dimension the features that are available on the CurseForge platform and that we can extract in an automated fashion. We end up with 33 features (characteristics) that we leverage to understand the differences between the characteristics of popular and unpopular Minecraft mods. 

Table~\ref{tab:metrics} shows an overview of the 33 features and their associated dimensions, along with their corresponding explanation and rationale. 
In addition, we normalized all features with the `\textit{numeric}' type in Table~\ref{tab:metrics} using a \code{log(1 + x)} transformation to reduce the bias caused by the outliers.

\begin{table}[]
\caption{Dimensions and their features describing the characteristics of popular and unpopular Minecraft mods.}
\centering
\label{tab:metrics}
\resizebox{\textwidth}{!}{
\begin{tabular}{p{1.3in}p{1.6in}p{1.8in}p{0.6in}p{1.8in}}
\toprule
\textbf{Dimension} & \textbf{Feature Name}& \textbf{Explanation} & \textbf{Type} & \textbf{Rationale} \\
\midrule
 Mod Category & \vtop{\hbox{\strut Number of categories}\hbox{\strut (\textit{num\_categories})}} & Total number of categories that a mod belongs to. A mod must belong to at least one category. & Numeric & Mods that offer a variety of categories can attract users with more options. \\
\hhline{~----}
 & \vtop{\hbox{\strut Miscellaneous}\hbox{\strut \textit{(is\_cat\_misc)}}} & Mods that do not belong to any of the existing categories. For example, the \textit{OpenBlocks} mod.\textsuperscript{1} & Boolean & Certain Minecraft mod categories in the CurseForge mod distribution platform may attract more users to the mod. \\
& \vtop{\hbox{\strut Food}\hbox{\strut \textit{(is\_cat\_food)}}} & Mods that provide changes to anything related to food in-game. For example, the \textit{AppleSkin} mod.\textsuperscript{2} & Boolean & \\
& \vtop{\hbox{\strut World generation}\hbox{\strut \textit{(is\_cat\_world\_gen)}}} & Mods that provide changes related to the world, such as new terrains.&Boolean& \\
& \vtop{\hbox{\strut Magic}\hbox{\strut \textit{(is\_cat\_magic)}}} & Mods that provide changes related to magic in the Minecraft game. For example, the \textit{Roots} mod.\textsuperscript{3}&Boolean & \\
& \vtop{\hbox{\strut API and library}\hbox{\strut \textit{(is\_cat\_library\_api)}}} & Mods that provide shared code for other mod developers to use. &Boolean& \\
& \vtop{\hbox{\strut Fabric}\hbox{\strut \textit{(is\_cat\_fabric)}}} & Mods that are created using the \textit{Fabric}\textsuperscript{4} modding toolchain. &Boolean & \\
& \vtop{\hbox{\strut Technology}\hbox{\strut \textit{(is\_cat\_technology)}}} & Mods that provide changes for any in-game technology. &Boolean & \\
& \vtop{\hbox{\strut Armor, tools, and weapons}\hbox{\strut \textit{(is\_cat\_armor\_weapons\_tools)}}} & Mods that provide changes to in-game armor, weapons, and tools. &Boolean& \\
& \vtop{\hbox{\strut Addons}\hbox{\strut \textit{(is\_cat\_addons)}}} & Mods that provide utilities for mod users to easily extend in-game features. &Boolean & \\
& \vtop{\hbox{\strut Adventure and RPG}\hbox{\strut \textit{(is\_cat\_adventure\_rpg)}}} & Mods that change the gameplay experience of the in-game adventure. &Boolean & \\
& \vtop{\hbox{\strut Server utility}\hbox{\strut \textit{(is\_cat\_server\_utility)}}} & Mods that provide changes to the server-side of the Minecraft game. &Boolean& \\
& \vtop{\hbox{\strut Redstone}\hbox{\strut \textit{(is\_cat\_redstone)}}} & Mods that are provide changes related to the redstone resource in the Minecraft game. &Boolean& \\
& \vtop{\hbox{\strut Map and information}\hbox{\strut \textit{(is\_cat\_map\_info)}}} & Mods that provide changes related to the location and information on items. &Boolean& \\
& \vtop{\hbox{\strut Storage}\hbox{\strut \textit{(is\_cat\_storage)}}} & Mods that provide mod users blocks and items, which improve the existing in-game storage. &Boolean & \\
& \vtop{\hbox{\strut Twitch integration}\hbox{\strut \textit{(is\_cat\_twitch\_integration)}}} & Mods that provide changes related to the interaction between the mod and the Twitch platform. &Boolean & \\
& \vtop{\hbox{\strut Cosmetic}\hbox{\strut \textit{(is\_cat\_cosmetic)}}} & Mods that provide changes to the texture and aesthetic of the in-game models. &Boolean & \\
\bottomrule
Mod Documentation & \vtop{\hbox{\strut Number of words in the}\hbox{\strut short description}\hbox{\strut \textit{(num\_words\_short\_desc)}}} & Number of words in the mod's preview description. & Numeric& The longer the description, the more likely that mod users will understand what the mod offers without downloading the mod. \\
& \vtop{\hbox{\strut Number of words in the}\hbox{\strut long description}\hbox{\strut \textit{(num\_words\_long\_desc)}}} & Number of words in the mod's main description.&Numeric & \\
\hhline{~----}
& \vtop{\hbox{\strut Mod wiki URL}\hbox{\strut \textit{(is\_mod\_wiki\_url)}}} & An external link with the documentation of a mod. &Boolean& The presence and quality of a mod's documentation can help other mod users understand how to utilize the mod to its full potential, which can give users a better experience. \\
\hhline{~----}
& \vtop{\hbox{\strut Number of images}\hbox{\strut \textit{(num\_images)}}} & Number of in-game screenshots that a mod has. &Numeric& In-game screenshots can help promote and visually explain the mod's functionalities, which may attract users, without trying the mod first. \\
\bottomrule
\end{tabular}
}
\end{table}

\begin{table}[]
\centering
\resizebox{\textwidth}{!}{
\begin{tabular}{p{1.3in}p{1.6in}p{1.8in}p{0.6in}p{1.8in}}
\toprule
Environmental Context of the Mod & \vtop{\hbox{\strut Latest number of incompatible} \hbox{\strut dependencies}\hbox{\strut \textit{(num\_incompatible\_dep)}}} & Number of incompatible dependencies that are in the latest release of the mod, which means that another mod is not compatible with the mod. &Numeric & Dependencies provide mods more functionalities, which could make the mod appeal to more users. \\
& \vtop{\hbox{\strut Latest number of tool}\hbox{\strut dependencies}\hbox{\strut \textit{(num\_tool\_dep)}}} & Number of words in the mod's main description.&Numeric & \\

& \vtop{\hbox{\strut Latest number of required}\hbox{\strut dependencies}\hbox{\strut \textit{(num\_required\_dep)}}} & Number of required dependencies that are in the latest release of the mod, which means another mod is required to make the mod function. &Numeric&\\

& \vtop{\hbox{\strut Latest number of embedded}\hbox{\strut library dependencies}\hbox{\strut \textit{(num\_embedded\_lib\_dep)}}} & Number of embedded library dependencies that are in the latest release of the mod, which provides shared code for the mod's development. &Numeric&\\

& \vtop{\hbox{\strut Latest number of optional}\hbox{\strut dependencies}\hbox{\strut \textit{(num\_optional\_dep)}}} & Number of optional dependencies that are in the latest release of the mod, which means the dependency adding a certain functionality can be switched on and off. &Numeric &\\\\
\hhline{~----}
& \vtop{\hbox{\strut Latest number of supported}\hbox{\strut Minecraft version(s)}\hbox{\strut \textit{(latest\_num\_mc\_versions)}}} & Number of Minecraft versions supported by the latest mod release, which corresponds to a specific version of the Minecraft game. Mods must support at least one Minecraft version. &Numeric & A larger number of supported versions could attract more users to a mod by providing more stability, and access to functionalities from different versions.\\

& \vtop{\hbox{\strut Latest number of supported}\hbox{\strut Java version(s)}\hbox{\strut \textit{(latest\_num\_java\_versions)}}} & Number of Java versions supported by the latest mod release. Mod developers can optionally provide this information. &Numeric & \\

& \vtop{\hbox{\strut Latest number of supported}\hbox{\strut Bukkit version(s)}\hbox{\strut \textit{(latest\_num\_bukkit\_versions)}}} & Number of Bukkit API\textsuperscript{5} versions supported by the latest mod release. The Bukkit API extends the multiplayer server of the Minecraft game for others to modify. Mod developers can optionally provide this information &Numeric & \\

\bottomrule 
Remuneration for the Mod& \vtop{\hbox{\strut PayPal URL}\hbox{\strut \textit{(is\_paypal\_url)}}} & An external link to PayPal for donations. &Boolean& Mod developers that ask for donations are more likely to be dedicated to modding, which can attract more users.\\

& \vtop{\hbox{\strut Patreon URL}\hbox{\strut \textit{(is\_patreon\_urls)}}} & An external link to Patreon for donations. &Boolean&\\

\bottomrule
Community Contribution for the Mod & \vtop{\hbox{\strut Source code URL}\hbox{\strut \textit{(is\_mod\_source\_code)}}} & An external link to the source code of a mod (e.g., Github). &Boolean & Mods that provide a link to their source code could invite more contributors, which could attract users with more content at a faster speed.\\
\hhline{~----}
& \vtop{\hbox{\strut Issue tracking URL}\hbox{\strut \textit{(is\_mod\_issues)}}} & An external link to an issue tracking system. &Boolean & Mods that provide a link to an issue tracking system could indicate to a user that a mod is more stable, which may attract them to the mod. \\

\bottomrule
\end{tabular}}
\begin{minipage}[t]{15cm}

	\begin{tablenotes}
 \small
	\item[1]\url{https://www.curseforge.com/minecraft/mc-mods/openblocks}
	\item[2]\url{https://www.curseforge.com/minecraft/mc-mods/appleskin}
	\item[3]\url{https://www.curseforge.com/minecraft/mc-mods/roots} \newline
	\item[4]\url{http://fabricmc.net/} \newline
	\item[5]\url{https://dev.bukkit.org/}
\end{tablenotes}
\end{minipage}
\end{table}

\section{Characteristics of Popular and Unpopular Minecraft Mods}
\label{sec:results}
In this section, we present the results of our empirical study of the characteristics of popular and unpopular Minecraft mods.

\subsection{RQ1: Do our studied dimensions have enough explanatory power to distinguish popular mods from unpopular ones?}

\label{sec:rq1}

\noindent\textbf{Motivation: }In this research question, we investigate how well each studied dimension of characteristics (i.e., features) of mods can individually distinguish the popular mods from unpopular ones. We also investigate how well can all the studied dimensions together distinguish popular mods from unpopular ones. Prior study~\cite{mobilechar} used similar dimensions to identify the characteristics that distinguish mobile apps with high ratings from the ones with low ratings. The results of this research question lay the foundation for further investigations of the characteristics of popular mods.

\noindent\textbf{Approach: }To investigate how well the individual dimensions can distinguish popular mods from unpopular ones (i.e., their explanatory power), we built a logistic regression model for each dimension in Table~\ref{tab:metrics}. We used logistic regression, instead of other complex techniques (e.g., a neural network) as logistic regression is transparent and interpretable~\cite{ruiz2008storms,molnar2018interpretable}. In particular, for each dimension's model, we used the features in a dimension as independent variables and whether the mod is popular as the dependent variable. We consider the given dimension to have significant explanatory power if the AUC of the model constructed with the dimension is greater than 0.5, which means that the dimension can distinguish popular from unpopular mods. The dimension that results in the largest AUC is deemed to have the most explanatory power and vice versa. We used the \code{glm} function\footnote{\url{https://www.rdocumentation.org/packages/stats/versions/3.6.1/topics/glm}} from the \code{stats} package\footnote{\url{https://www.rdocumentation.org/packages/stats/versions/3.6.1}} to create the logistic regression models. 

To validate the performance of our built models, we performed 100 out-of-sample bootstrap iterations to compute the AUC (Area Under the receiver operator characteristics Curve) for each model. Prior study~\cite{tantithamthavorn2016empirical} showed that the out-of-sample bootstrap technique had the best balance between the bias and variance of estimates. The out-of-sample bootstrap technique randomly samples data with replacement for \code{n} iterations. The sampled data in an iteration is used as the training set for that iteration, while the data that was not sampled in that iteration is used as the testing set for that iteration. We then trained a model with the training set and calculated the AUC of the model with the testing set for each iteration.

In addition, to investigate how well all studied dimensions combined can distinguish popular mods from unpopular mods, we built a logistic regression model using all 33 features from the 5 dimensions in Table~\ref{tab:metrics}. We evaluated the performance of this combined model using the same aforementioned process of computing the AUC of the model with 100 out-of-sample bootstrap iterations.

\begin{figure*}[!t]
	\center
\includegraphics[width=0.7\textwidth]{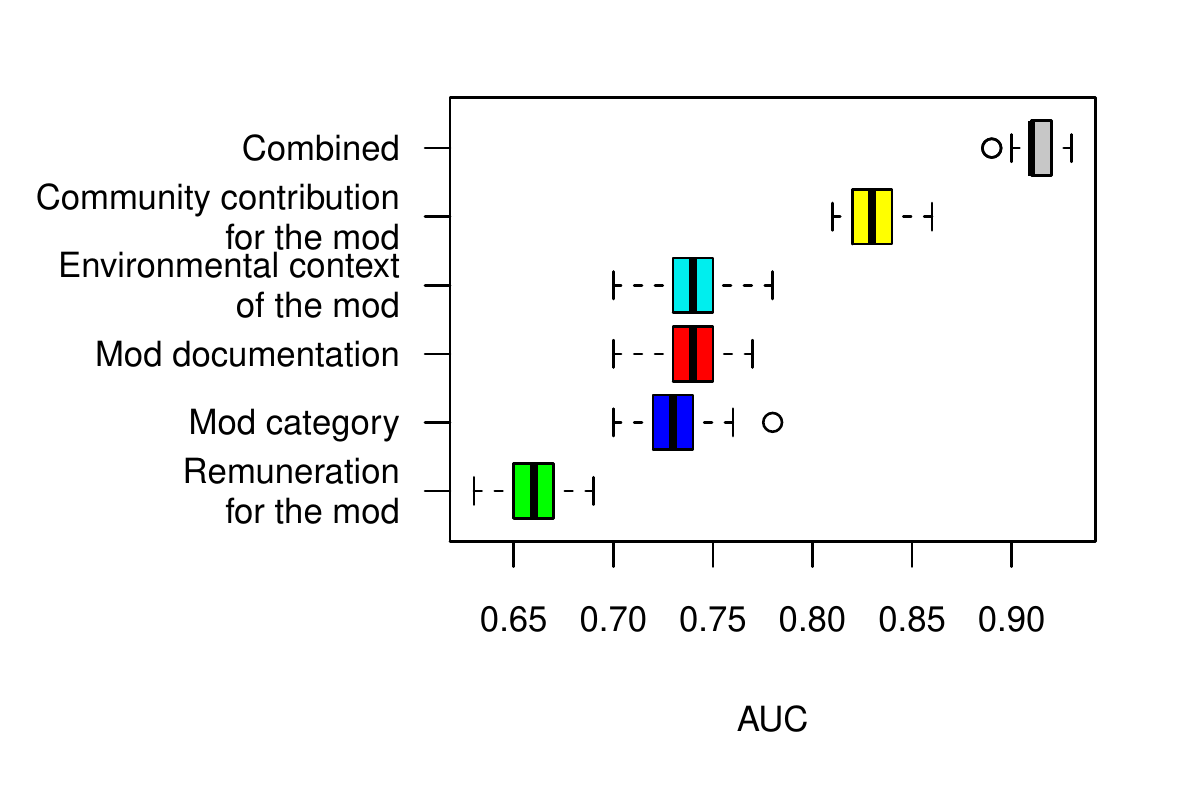}
	\caption{The distribution of the AUCs of models constructed with an individually studied dimension, and with all studied dimensions combined. The different colors represent the statistically different ranks given by the Scott-Knott effect size difference test. The distributions are sorted by their ranks (presented in ascending order from left to right with remuneration for the mod having the lowest rank) from the Scott-Knott effect size difference test.}
	\label{fig:rq1_boxplot_aucs}
\end{figure*}

Furthermore, we used the Scott-Knott effect size difference test to statistically sort and rank the distributions of the AUCs of all studied dimensions~\cite{tantithamthavorn2016empirical}. We used the \code{sk\_esd} function\footnote{\url{https://www.rdocumentation.org/packages/ScottKnottESD/versions/1.2.2/topics/\%22sk\_esd\%22}} from the \code{ScottKnottESD} package\footnote{\url{https://www.rdocumentation.org/packages/ScottKnottESD/versions/1.2.2}} for the Scott-Knott effect size difference test.

\noindent\textbf{Findings: }\textbf{Each studied dimension has significant explanatory power to individually identify popular mods.} Figure~\ref{fig:rq1_boxplot_aucs} shows the distribution of AUCs per studied dimension. The lowest median AUC among the studied dimensions was 0.66, implying that every dimension has significant explanatory power (i.e., the model has an AUC $>$ 0.5) in distinguishing popular mods from unpopular ones. In addition, the Scott-Knott effect size difference test shows a statistical significant difference between each studied dimensions, with non-negligible effect sizes. Among the studied dimensions, the community contribution for the mod dimension is ranked as having the largest explanatory power, whereas the remuneration for the mod dimension is ranked as having the lowest explanatory power.

\textbf{The combined model has a larger explanatory power than each of the studied dimension individually.} Figure~\ref{fig:rq1_boxplot_aucs} shows the distribution of AUCs of the \emph{combined model} that combines all studied dimensions together. The combined model has the largest median AUC of 0.91, outperforming every one of the studied dimensions on their own. The Scott-Knott effect size difference test confirms that the combined model has the highest ranking in explanatory power compared to the individual studied dimensions. 

In addition, Figure~\ref{fig:rq1_boxplot_aucs} shows that the combined model has a 10\% higher median AUC than the community contribution for the mod dimension (the dimension with the highest explanatory power among the studied dimensions), and a 38\% higher median AUC than the remuneration for the mod dimension (the dimension with the lowest explanatory power among the studied dimensions). Prior study~\cite{mobilechar} also observed that a combined model with all the dimensions has a larger explanatory power than models with individual dimensions in the context of distinguishing mobile apps with high ratings from mobile apps with low ratings. 

\hypobox{Each studied dimension of characteristics of a mod has significant explanatory power in distinguishing popular from unpopular mods. Among the studied dimensions, the community contribution for the mod dimension has the largest explanatory power. However, our combined model which uses all the features across the five dimensions outperforms the best model using individual dimension by 10\% (median).} 

\subsection{RQ2: Which features best characterize a popular mod?}
\label{sec:rq2}

\noindent\textbf{Motivation: } In this research question, we investigate which mod features can best characterize popular mods. The results of RQ1 show that the studied dimensions have a strong explanatory power for the popularity of a mod. In this RQ, we further investigate the characteristics of popular mods at the feature-level across 33 features and 5 dimensions to systematically quantify the association between the studied features and the number of downloads for a mod.

\begin{figure*}[!t]
	\center
\includegraphics[width=\textwidth]{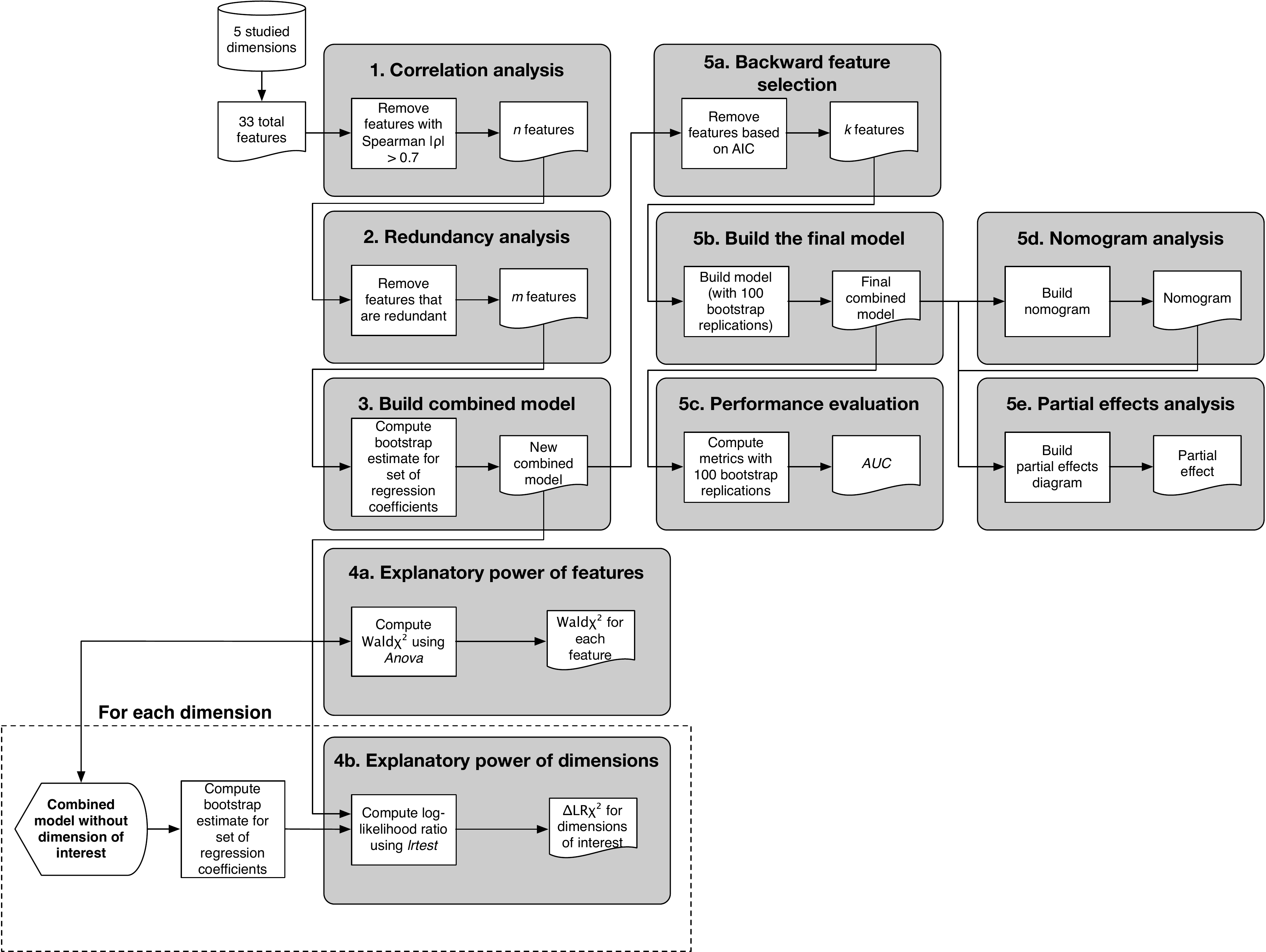}
	\caption{An overview of the process that we used to build, evaluate and analyze the combined model.}
	\label{fig:rq2c_experiment_setup}
\end{figure*}

\noindent\textbf{Approach: }To investigate which features can best characterize popular mods, in this research question we focus on analyzing the combined model with all dimensions of features, as RQ1 shows that the combined model has the most explanatory power for mod popularity. 

Figure~\ref{fig:rq2c_experiment_setup} shows an overview of our approach to construct, evaluate and analyze the combined model. Below we explain each step in detail: 
\begin{description}[leftmargin=*]
\item1. \textbf{Correlation analysis.} We performed correlation analysis to reduce collinearity between the features before we built the models, since correlated features can affect the interpretation of the model~\cite{midi2010collinearity,mcintosh2016empirical}. We used the \code{varclus} function\footnote{\url{https://www.rdocumentation.org/packages/Hmisc/versions/4.2-0/topics/varclus}} from the \code{Hmisc} package\footnote{\url{https://cran.r-project.org/web/packages/Hmisc/index.html}} in \code{R} to filter out highly correlated features. We calculated \textit{Spearman's correlation coefficients} among the studied features. We consider a pair of features with a \textit{Spearman correlation coefficient} $>=$ 0.7 as highly correlated. We did not observe high correlations among our studied features.

\item2. \textbf{Redundancy analysis.} Before building the models, we also performed redundancy analysis to eliminate redundant features that can interfere with the relationship between the independent variables (i.e., features), which in turn may distort the relationship the independent variables have with the dependent variable (i.e., popularity)~\cite{mcintosh2016empirical}. We used the \code{redun} function\footnote{\url{https://www.rdocumentation.org/packages/Hmisc/versions/4.2-0/topics/redun}} from the \code{Hmisc} package in \code{R} to filter out features that can be linearly predicted by other features. We removed the `\textit{number of categories}' feature as it is redundant, leaving 32 features for the remainder of the study.

\item3. \textbf{Building the combined model.} We used all the remaining features after step 2 to build a logistic regression model. However, the model's regression coefficients could vary or be estimated incorrectly based on the sample of data and the underlying assumptions~\cite{fox2002bootstrapping}. Hence, to avoid biasing the estimated regression coefficients, we used the \code{bootcov} function from the \code{rms} package using 100 bootstrap iterations to adjust the regression coefficients with bootstrap estimates, to ensure the non-arbitrariness of the estimated regressions co-efficients in the combined model~\cite{harrell1984regression,harrell2001introduction}.

\item4a. \textbf{Explanatory power of features.} We used Wald's $\chi^2$ to measure the explanatory power of the features in the model from step 3. The larger the Wald $\chi^2$, the larger the explanatory power of the feature~\cite{harrell1984regression}. 
Prior study~\cite{reviewdynamics} used the same approach to compute the explanatory power of features. We computed the Wald $\chi^2$ with the \code{Anova} function\footnote{\url{https://www.rdocumentation.org/packages/car/versions/3.0-3/topics/Anova}} from the \code{car} package\footnote{\url{https://www.rdocumentation.org/packages/car/versions/3.0-3}} in \code{R} using the parameter \textit{test.statistic=`Wald'}. Table~\ref{tab:statresults} shows the explanatory power of each feature (Wald $\chi^2$).

\item4b. \textbf{Explanatory power of dimensions.} 
Though in RQ1, we observed that each dimension of features of a mod has explanatory power, we are uncertain of the unique explanatory power each of them contains in relation to the other dimensions. Understanding the unique explanatory power of each dimension is critical to assert which of these dimensions matter the most for characterizing the popularity of a mod. For example, from Figure~\ref{fig:rq1_boxplot_aucs} we observe that the environmental context of the mod and mod documentation dimensions by themselves can explain the popularity of a mod with a median AUC of 0.74. However, we are uncertain of how much unique power each of these dimensions contribute to the model built on all the studied dimensions, which had a median AUC of 0.92. 

\smallskip

Therefore, we conducted a chunk test on each of the studied dimensions in the combined model from step 3, to quantify the explanatory power of each studied dimension~\cite{chunktest,mcintosh2016empirical}. For each of the studied dimensions (given in Table~\ref{tab:metrics}), the chunk test estimates the difference in goodness of fit (by computing the difference in log-likelihood) between the full model (i.e., the combined model from step 3) and the combined model that was built without one studied dimension (whose explanatory power we are computing). The chunk test reports a Chi-square value ($\Delta$LR$\chi^2$) (which is the difference in log-likelihood compared to the Chi-squared distribution) and a p-value. The Chi-squared value quantifies the unique explanatory power that was lost due to the removal of the given dimension (in relation to the other dimensions) and a lower p-value ($<=0.05$) signifies the dimension's significance.

We used the \code{lrtest} function\footnote{\url{https://www.rdocumentation.org/packages/lmtest/versions/0.9-37/topics/lrtest}} from the \code{lmtest} package\footnote{\url{https://www.rdocumentation.org/packages/lmtest/versions/0.9-37}} in \code{R} to conduct the chunk test. Table~\ref{tab:statresults} shows the explanatory power of each dimension ($\Delta$LR$\chi^2$).

\begin{table}[]
\centering
\caption{An overview of the statistics of each dimension and its features. The larger the $\Delta$LR$\chi^2$, the larger the role of a studied dimension. Similarly, the larger the Wald $\chi^2$, the larger the explanatory power of a feature in the combined model (The percentages and p-value are rounded to two decimal places). The feature is statistically significant if the p-value $<=$ 0.05. Sorted by the Wald $\chi^2$ per studied dimension.}
\begin{tabular}{lrr}
\toprule
 & \textbf{Wald} $\chi^2$ (\%)& \textbf{P-value} \\
 \toprule
 \vtop{\hbox{\strut \textbf{Mod Category} } \hbox{\strut \textbf{($\Delta$LR$\chi^2$: 23.51\%)}}}
 & & 0.00\\ 
 \hhline{---}
 Fabric & 20.54 & $<$ 0.01\\
 Armor, tools, and weapons & 7.92 & $<$ 0.01\\
 Addons & 2.22 & $<$ 0.01\\
 Food & 2.17 & $<$ 0.01\\
 World generation & 1.64 & $<$ 0.01\\
 API and library & 1.47 & $<$ 0.01\\
 Miscellaneous & 1.39 & $<$ 0.01\\
 Server utility & 1.17 & $<$ 0.01\\
 Storage & 0.79 & 0.02\\
 Redstone & 0.50 & 0.06\\
 Adventure and RPG & 0.43 & 0.08\\ 
 Cosmetic & 0.27 & 0.17\\ 
 Technology & 0.10 & 0.39\\ 
 Map and information & 0.02 & 0.73\\ 
 


 Magic & 0.01 & 0.74\\ 









 Twitch integration & 0.00 & 0.97\\ 

\hhline{---}
 \vtop{\hbox{\strut \textbf{Mod Documentation}} \hbox{\strut \textbf{($\Delta$LR$\chi^2$: 13.03\%)}}}
 & & 0.00 \\
\hhline{---}
 Number of words in the long description & 5.18 & $<$ 0.01\\
 Number of images & 2.06 & $<$ 0.01\\
 Mod wiki URL & 1.72 & $<$ 0.01\\
 Number of words in the short description & 0.46 & 0.07\\



\hhline{---}
 \vtop{\hbox{\strut \textbf{Environmental Context of the Mod}} \hbox{\strut \textbf{($\Delta$LR$\chi^2$: 18.63\%)}}}
 & & 0.00\\
\hhline{---}
 Latest number of Bukkit versions& 15.45 & $<$ 0.01\\
 Latest number of required dependencies& 9.27 & $<$ 0.01\\
 Latest number of optional dependencies& 0.81 &0.02\\
 Latest number of Minecraft versions& 0.49 & 0.07\\
 Latest number of Java versions& 0.06 & 0.53\\ 
 Latest number of tool dependencies& 0.00 & 0.85\\ 
 Latest number of incompatible dependencies& 0.00 & 0.86\\ 


 Latest number of embedded library dependencies& 0.00 & 0.98\\ 

 \hhline{---}
 \vtop{\hbox{\strut \textbf{Remuneration for the Mod}} \hbox{\strut \textbf{($\Delta$LR$\chi^2$: 10.42\%)}}}
 & & $<$ 0.01\\
 \hhline{---}
 Paypal URL& 6.21 & $<$ 0.01\\
 Patreon URL& 1.56 & $<$ 0.01\\
 \hhline{---}
 \vtop{\hbox{\strut \textbf{Community Contribution for the Mod}} \hbox{\strut \textbf{($\Delta$LR$\chi^2$: 34.41\%)}}}
 & & 0.00\\
 \hhline{---}
 Issues URL& 11.21 & $<$ 0.01\\
 Source code URL& 4.86 & $<$ 0.01\\
 \bottomrule
 \textbf{Total} & 100.00 & \\
\bottomrule
\end{tabular}
\label{tab:statresults}
\end{table}

\item{5a. \textbf{Backward feature selection. }}We do backward feature selection to ensure the parsimony of the constructed model, as suggested by Harrell et al.~\cite{harrell1984regression}. For instance, if a model contains a large number of independent features, the model becomes too complex to draw explanations. Hence, Harrell et al.~\cite{harrell1984regression} suggests using backward feature selection when the goal of the model is to interpret it. 
We used the \code{fastbw} function\footnote{\url{https://www.rdocumentation.org/packages/rms/versions/5.1-3.1/topics/fastbw}} from the \code{rms} package in \code{R} to perform a backward elimination of features. The~\code{fastbw} function takes the model that was constructed on all the features (32) and eliminates the features that do not significantly contribute to reducing the AIC of the model.  
We removed 14 of the 32 features (44\%) using the \code{fastbw} function. In result, we obtained a new combined model with 18 features.

\item{5b. \textbf{Build the final model. }} With the reduced feature set from step 5a, we reconstructed the final combined model. Similar to step 3, we adjusted the regression coefficients with the bootstrap estimate, as outlined by Harrell et al.~\cite{harrell1984regression}.

\item{5c. \textbf{Performance evaluation. }}
To demonstrate the quality of the constructed model from 5b, we calculated the AUC of the model using 100 out-of sample bootstrap iterations to evaluate the performance of the model.

\item{5d. \textbf{Nomogram analysis. }}We used the final combined model from step 5b to create and analyze a nomogram using the \code{nomogram} function\footnote{\url{https://www.rdocumentation.org/packages/rms/versions/5.1-3.1/topics/nomogram}} from the \code{rms} package in \code{R}, which provides a way to measure the explanatory power of each feature in distinguishing popular from unpopular mods. A nomogram provides a graphical visualization of the parsimonious logistic regression model that we built in step 5b. Although the Wald $\chi^2$ can provide insight into the explanatory power of each feature in the combined model, the nomogram provides us with an exact interpretation on how the variation in each feature affects the outcome probability. For instance, while the Wald $\chi^2$ may indicate that the number of words in the long description of a mod is important, the Wald $\chi^2$ does not provide insights on how the exact number of words in the long description contribute to the explanatory power in distinguishing popular from unpopular mods. Furthermore, the Wald $\chi^2$ does not show if a certain feature has a positive or negative role in distinguishing popular from unpopular mods, whereas the nomogram does. For instance, if for a given mod, the feature ``\textit{latest\_num\_bukkit\_versions}'' is 0, then it has a positive role in distinguishing popular from unpopular mods. Several prior studies~\cite{shariat2009use,chun2007critical} showed that nomograms are one of the most accurate discriminatory tools in interpreting a logistic regression model. Hence, we constructed a nomogram to observe the exact role of features in classifying if a given mod is either popular or unpopular. Another key difference between the Wald $\chi^2$ and nomogram is that the nomogram can show the contribution of each feature towards the outcome probability for each of the studied mods, whereas the Wald $\chi^2$ only shows the overall contribution (which is not specific to each mod). Figure~\ref{fig:rq2_nomogram_overall} shows the results of the nomogram analysis.

\item{5e. \textbf{Partial effects analysis. }}We used the final combined model from step 5b and the nomogram analysis from step 5d to create partial effects plots, which show how different values in numeric features with respect to another feature held constant at the median for numeric features and at the mode for boolean features, contributes the outcome probability. Hence, the partial effects analysis provides a deeper explanation of how the variation in certain features can contribute to the probability of a mod being popular or unpopular.
\end{description}

In addition, to measure if two distributions are significantly different, we used the Wilcoxon tests. The Wilcoxon signed-rank test is a paired and non-parametric statistical test, whereas the Wilcoxon rank-sum test is an unpaired and non-parametric statistical test, where the null hypothesis indicates that it is equally likely that a randomly selected value from one sample will be less than or greater than a randomly selected value from a second sample~\cite{wilcoxon1945individual}. If the p-value of the used Wilcoxon test on the two distributions is less than 0.05, we reject the null hypothesis, and conclude that the two distributions are significantly different. In addition, to calculate the magnitude of the difference we calculate the Cliff's delta \textit{d} effect size~\cite{long2003ordinal}, with the following thresholds~\cite{romano2006exploring}: 

\begin{equation*}
 \text{Effect size}=\begin{cases}
 \textit{negligible(N)}, & \text{if $|d| \leq 0.147$}.\\
 \textit{small(S)}, & \text{if $0.147 < |d| \leq 0.33$}.\\
 \textit{medium(M)}, &\text{if $0.33 < |d| \leq 0.474$}.\\
 \textit{large(L)}, & \text{if $0.474 < |d| \leq 1$}. \\
 \end{cases}
\end{equation*}

\begin{figure*}[!t]
	\center
\includegraphics[width=1.0\textwidth]{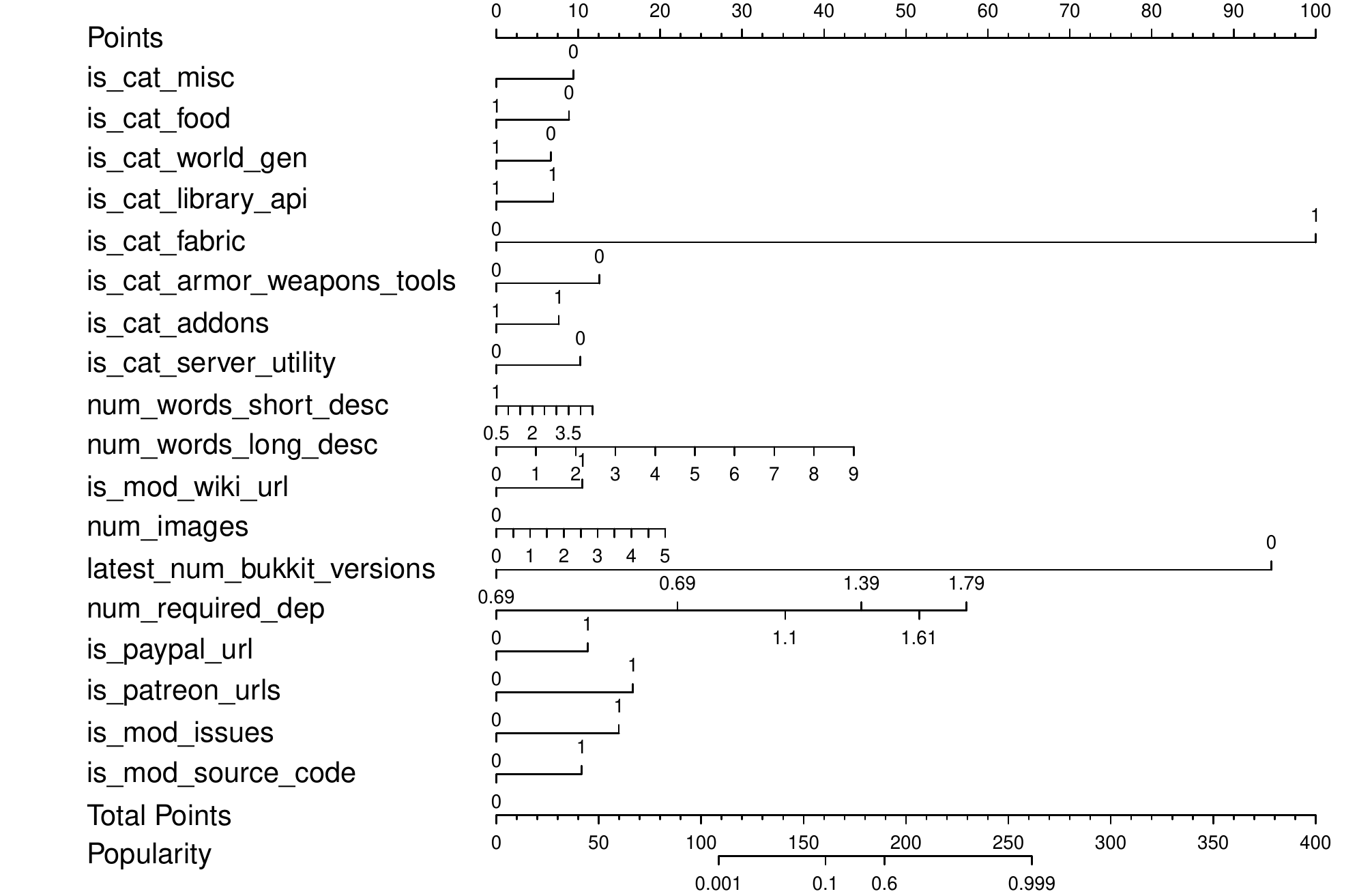}
	\caption{The nomogram visualizes the role of each feature in distinguishing a mod's popularity. The line against each feature in the figure, varies between the range of values for that given feature. The ``\textit{points}'' line at the top of the figure, is used to calculate the magnitude of contribution that each feature has and ``\textit{Total Points}'' at the bottom of the figure gives the total points generated by all the features for a given instance (i.e., for a given mod). For instance, if for a given mod, the feature ``\textit{is\_cat\_fabric}'' has a value of 1, then it contributes 100 points. Finally, the line against ``\textit{Popularity}'' shows the probability of a mod to be classified as a popular mod according to the total number of points (which is computed by summing up all the individual points contributed by each feature). For instance, if all the features for a given mod contribute a total of 260 points, then the probability of that mod to be classified as popular by our explanatory model is 99\% and similarly, if the total points given by all the features for a particular mod is less than 110, then that mod will be classified as not popular. Also, the model used to generate this nomogram achieved a median AUC of 0.92 on 100 out-of-sample bootstrap iterations.}

	\label{fig:rq2_nomogram_overall}
\end{figure*}

\noindent\textbf{Findings: }
\begin{figure}[!htbp]
	\centering
	\subfloat[Number of words in the long description w.r.t. a mod wiki URL.]{
		\includegraphics[width=.5\textwidth]{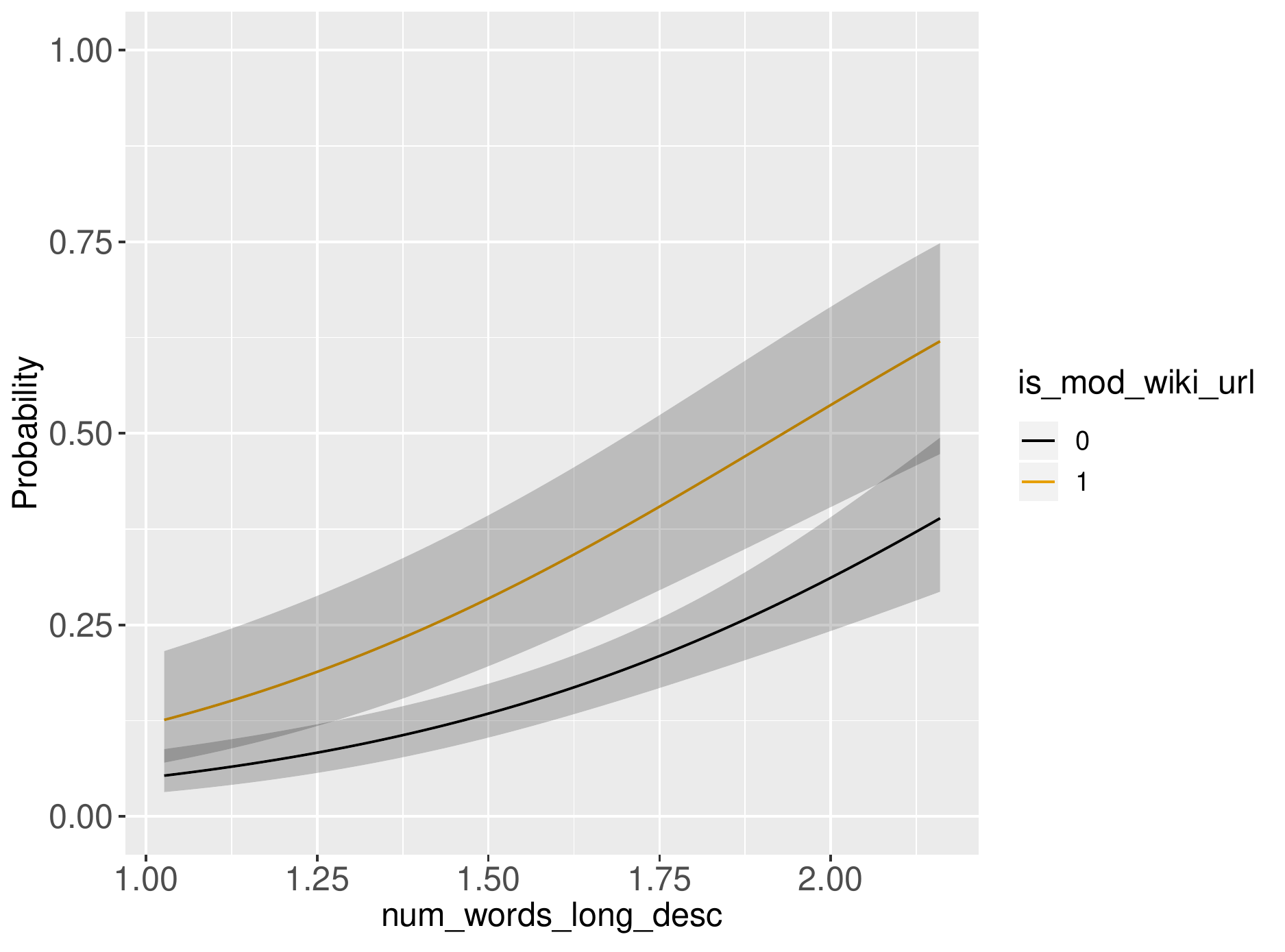}
		\label{fig:longdesc_wiki}
	}
	\subfloat[Number of images w.r.t. a PayPal URL.]{
		\includegraphics[width=.5\textwidth]{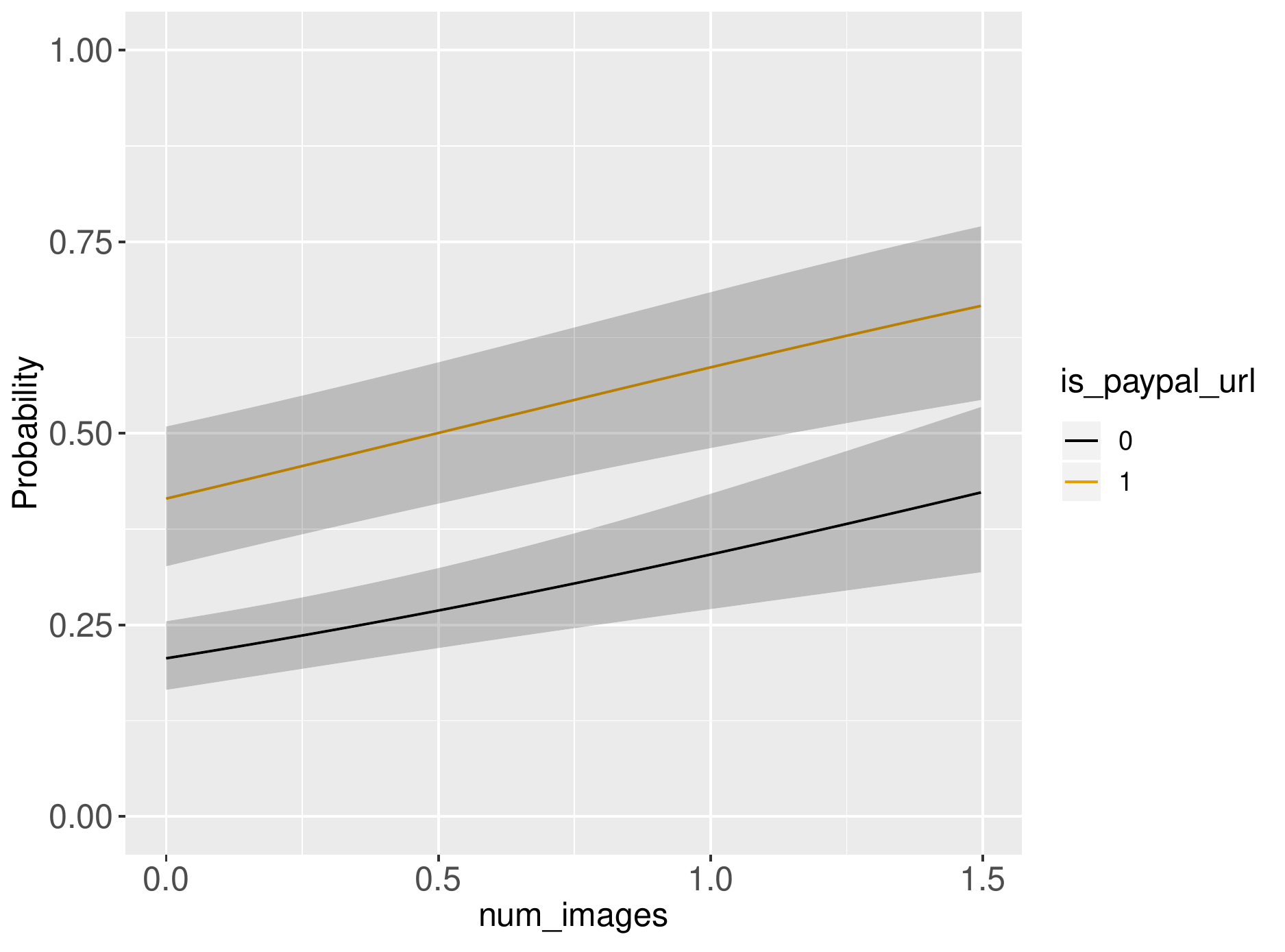}
		\label{fig:images_paypal}
	}\\
	\subfloat[Number of images w.r.t. a mod wiki URL.]{
		\includegraphics[width=.5\textwidth]{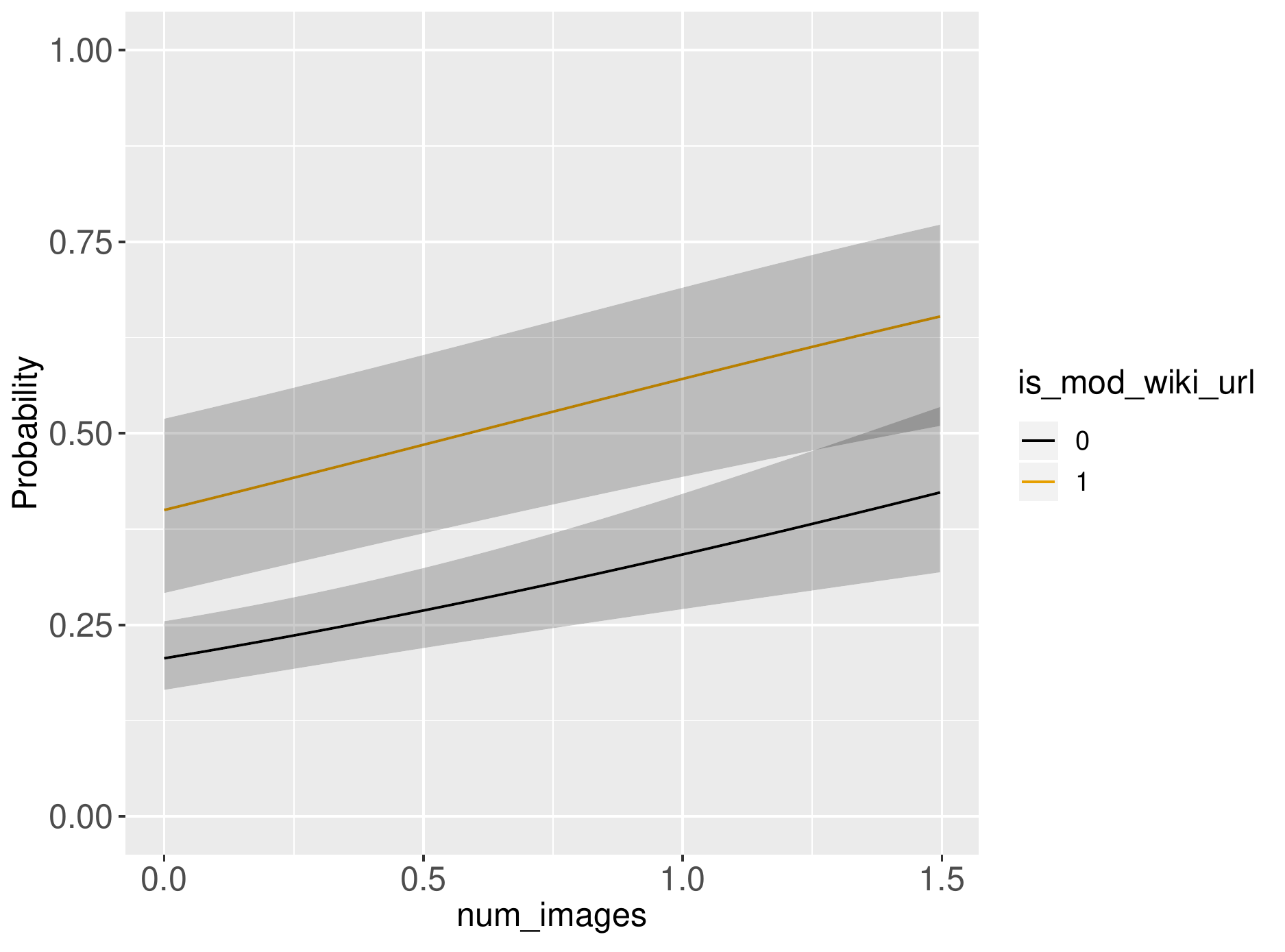}
		\label{fig:numimages_wiki}
	}
	\subfloat[Number of words in the long description w.r.t. an issue tracking URL.]{
		\includegraphics[width=.5\textwidth]{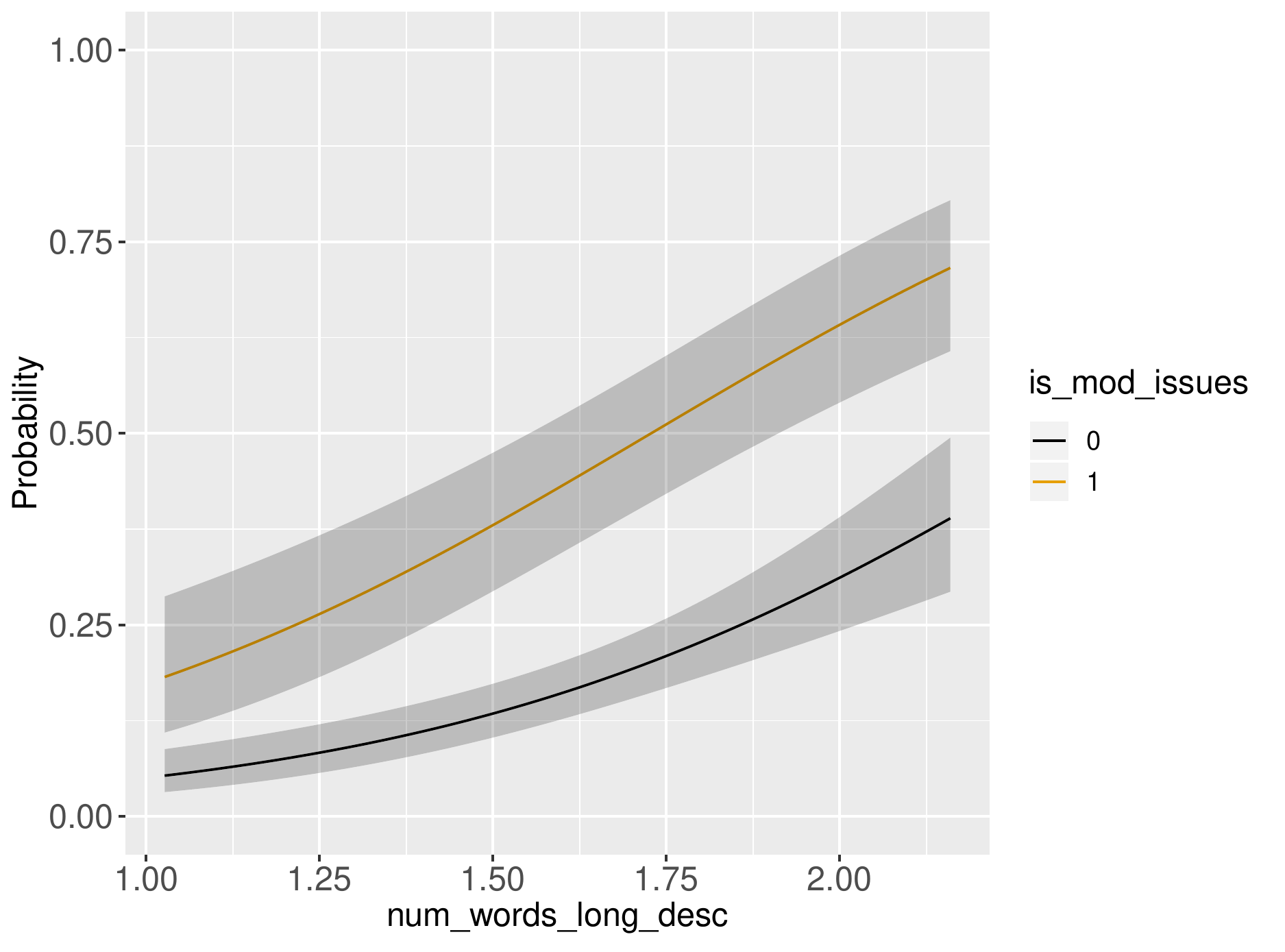}
		\label{fig:ismodissuesvsdesc}
	}\\
	\caption{The impact of features on the outcome probability when another feature is held constant (features are held constant at the median for numeric features and at the mode for boolean features). The grey area shows a confidence interval at 95\%.
	}
	\label{fig:partial_effects}
\end{figure}
\textbf{Mods that simplify mod development are a popular type of mods}. Figure~\ref{fig:rq2_nomogram_overall} shows that mods that belong to the \textit{``fabric''}, \textit{``addons''}, and \textit{``API and library''} categories tend to be among the most popular mods.
We further investigated the mods under each category and observed that all of the 16 collected \textit{``fabric''} mods are popular mods, 73.3\% of the studied \textit{``addons''} mods are popular mods, and 71.1\% of the studied \textit{``API and library''} category mods are popular mods. Mods of the \textit{``fabric''} category are created using the \textit{``fabric''} mod development toolchain, which offers a mod ecosystem that makes updating a mod simpler and provides modularity of the code~\cite{fabricapi}. 
Mods of the \textit{``API and library''} category can be leveraged by others and mod developers to make mod development simpler. In addition, mods of the \textit{``addons''} category, such as the TOP Addons mod, add support to and extend other mods\footnote{\url{https://www.curseforge.com/minecraft/mc-mods/top-addons}}. 

Finally, the \textit{``miscellaneous''}, \textit{``food''}, \textit{``world generation''}, \textit{``armor tools weapons''}, and \textit{``server utility''} mod categories are more related to unpopular mods. 

\textbf{Over 70\% of the studied popular mods include a source code URL and/or issue tracking URL}, as shown in Figure~\ref{fig:rq2_nomogram_overall}. 
We investigated the studied popular mods and observed that 77\% of the popular mods have an issue tracking URL, and 71\% of the popular mods have a source code URL. In addition, Figure~\ref{fig:ismodissuesvsdesc} shows that the presence of an issue tracking URL with at least about 145 words in the mod's main description increases the probability of distinguishing popular from unpopular mods.

Furthermore, from Table~\ref{tab:statresults}, we observe that the community contribution dimension (which captures the presence/absence of source code URL and/or an issue tracking URL) has the highest explanatory power (34.4\%) among all the other studied dimensions. Even though other individual features contribute towards characterizing the popularity of a mod, community contribution dimension as a whole is more important. 

\textbf{Popular mods have longer descriptions than unpopular mods.} The descriptions of popular mods have a median of 161.5 words, whereas the descriptions of unpopular mods have a median of 75 words. The Wilcoxon rank-sum test confirms that the number of words in the description of popular mods and unpopular mods is statistically significantly different, with a medium Cliff's delta effect size. 
In Figure~\ref{fig:longdesc_wiki}, we held the mod wiki URL at a constant against the number of words in the description because if a mod developer is willing to provide external documentation, they could be more willing to make an effort into providing a richer description for the mod. Prior work~\cite{mobilechar} showed that high-rated mobile apps had significantly longer app descriptions, which is consistent with our results.

In addition, Figure~\ref{fig:rq2_nomogram_overall} shows that popular mods have more images and a wiki URL. 
Therefore we posit that mod developers who make an effort to provide external documentation are likely to further explain how the mod works visually to users by presenting in-game screenshots, and Figure~\ref{fig:numimages_wiki} confirms this observation. Prior work~\cite{mobilechar} observed that the number of images is one of the top three influential factors in determining that a mobile app will be high-rated, which is consistent with the results of our study of mods.

Finally, the number of words in the description, the number of images, and having a wiki URL are all features that are related to the mod documentation dimension, and all of them have a positive relationship with mod popularity.

\textbf{Popular mods typically accepted donations and tended to be more active (i.e., they have more releases and comments).} Figure~\ref{fig:rq2_nomogram_overall} and~\ref{fig:images_paypal} show that popular mods often have a Paypal URL or Patreon URL. Mods with a PayPal URL have a median of 13 mod releases, whereas mods without a PayPal URL have a median of 2 mod releases; mods with a Patreon URL had a median of 21 mod releases, whereas mods without a Patreon URL had a median of 3 mod releases. The Wilcoxon rank-sum test confirms that the differences in the number of mod releases between mods with and without a PayPal URL or Patreon URL are both statistically significant, with a medium Cliff's delta effect size for a PayPal URL and a large Cliff's delta effect size for a Patreon URL. 

Furthermore, mods with a Patreon URL have a median of 25 comments per mod, while mods without a Patreon URL have a median of 1 comment per mod. The Wilcoxon rank-sum test confirms a statistically significant difference in the number of comments between mods with and without a Patreon URL, with a small Cliff's delta effect size. 

In total, we observed that 88 mod developers advertise their Patreon URL on their mods' pages. We manually investigated the motivation of them accepting donations by looking at each of their Patreon profiles. 14\% of these mod developers created a Patreon to support their living (e.g., pay bills), 32\% of them created a Patreon for fun and did not expect profit, 32\% of them created a Patreon to obtain motivation in continuously releasing new content (e.g., faster release of content), and 23\% of them either closed or did not finish setting up their Patreon profile. 

We further investigated the release frequency of mods (with more than 1 mod release) that are created by the 32\% of mod developers who use Patreon for motivation to release new content. 

However, the Wilcoxon rank-sum test shows no statistically significant difference in the release frequency between mods that are created by mod developers that accept donations for motivation to mod (a median mod release frequency of every 6 days) and mods that are created by other mod developers (a median mod release frequency of 7 days). The Wilcoxon rank-sum test did show a statistically significant difference in the number of mod releases between mods that are created by mod developers that accept donations to mod (a median number of 23 mod releases) and mods that are created by other mod developers (a median number of 11 mod releases), with a medium Cliff's delta effect size. Hence, mod developers who accept donations as a motivation to create mods do produce a larger number of mods than other mod developers (though not necessarily more popular mods). However, their release frequency is similar to the mod developers who do not accept donations as a motivation.

Interestingly, \textit{LexManos}\footnote{\url{https://www.patreon.com/lexmanos}} received the most donations at \$2,157 per month. LexManos is the creator and primary developer of the popular Minecraft Forge API~\cite{mostmodsrelyonforge}, which is a mod loader API that is required to run most Minecraft mods. However, other mod developers who have a valid Patreon URL only generate a median of \$4 per month.

\hypobox{18 of the 33 (54.5\%) studied features have a role in distinguishing popular mods from unpopular ones. Popular mods tend to promote community contributions with a source code URL and an issue tracking URL, and have a richer mod description.}

\section{Threats to Validity}
\label{sec:threats}
This section outlines the threats to the validity of our findings.

\subsection{Internal Validity}

A threat to the internal validity of our study is that we only studied the top and bottom 20\% of the mods (based on their number of downloads). However, the top and bottom 20\% of the mods ensures that there is a clear distinction between popular and unpopular mods, as mods having close to the median number of total downloads can belong to either one. Such approach is also used in prior study~\cite{mobilechar}.

Another threat to the internal validity of our study is that we only focused on the mods that were created between 2014 and 2016. However, such restriction is necessary to reduce the bias introduced by the extreme short or long lifetime of a mod.

An additional internal threat to validity is that we do not cover all the possible features that are related to mods. However, we conduct a first study to understand the characteristics of popular and unpopular mods specific to a particular game (Minecraft) and we encourage future work to explore additional features and dimensions. 

For example, Minecraft has been used as a sandbox for a plethora of activities, for example, in the education sector. Therefore, the educational value of a mod might potentially be an important confounder in determining the popularity of a mod in addition to the features that we observe in our study. We suggest that future studies investigate how the other latent functional and educational aspects of Minecraft modding affect its popularity using statistical procedures that are similar to the ones that are outlined in our study.

Finally, it is important to realize that mod developers of the CurseForge mod distribution platform could at anytime change the name of their mod, remove mod developers or delete the mod. As a result, some older mods or mod developers may not exist at the time of our data collection. Future studies should investigate the life cycle of mods and mod developers on the CurseForge mod distribution platform.

\subsection{External Validity}
A threat to the external validity of our study is that we only studied mods from the CurseForge mod distribution platform. However, the CurseForge mod distribution platform has the largest number of mods out of other mod distribution platforms, as shown in Section~\ref{sec:bg}. Furthermore, we clearly document the data collection and the statistical approach that we use to arrive at the characteristics of popular game mods in the CurseForge platform. Therefore, our approach could be replicated by other future studies that seek to investigate the characteristics of popular and unpopular mods across different mod distribution platforms (such as the Nexus mods platform). Another threat to the external validity of our study is that we only studied mods for the Minecraft game. Although the Minecraft game is one of the best selling games in 2019, and hosts one of most active and largest modding communities, our results may or may not generalize across mods developed for a different game. Therefore, future studies should use our outlined approach compare our results with mods of different games.

\section{Conclusion}
\label{sec:conclusion}

An active modding community not only helps game developers meet the growing and changing needs of their gamer base, but also leads to a better overall gaming experience. In this paper, we studied the characteristics of popular mods with a large number of downloads by analyzing 2,228 Minecraft mods from the CurseForge mod distribution platform, along 5 dimensions of characteristics for a mod: mod category, mod documentation, environmental context of the mod, remuneration for the mod, and community contribution for the mod. We firstly verified that the studied dimensions have significant explanatory power in distinguishing popular from unpopular mods. Then, we investigated the contribution of each of the 33 features across these 5 dimensions of mod characteristics on the popularity of a mod. The most important findings of our paper are:

\begin{enumerate}
\item The community contribution for the mod dimension has the strongest explanatory power of the popularity of mods. Popular mods tend to promote community contribution with a source code URL and an issue tracking URL.
\item Simplifying the mod development is positively correlated with mod popularity. 
\item Popular mods tend to have a high quality description.

\end{enumerate}

Based on our findings, we suggest future work to further investigate the impact of the features that distinguish popular mods, to eventually come with recommendations that assist mod developers in improving the popularity of their mods.

\bibliographystyle{spbasic}      

\bibliography{ms}

\end{document}